\documentclass{appolb}
\usepackage{graphicx}
\usepackage[T1]{fontenc}
\usepackage[utf8]{inputenc}

\begin{document}
\title{Pre-equilibrium effects of the hot nuclei de-excitation via GDR emission - theoretical approach
\thanks{Presented at XXIV Nuclear Physics Workshop
20-24 September 2017
Kazimierz Dolny, Poland}}%
\author{K. Mazurek, M. Ciema\l{}a, M. Kmiecik, A. Maj
\address{IFJ PAN, PL-31342 Krakow, Poland}\\
 D. Lacroix
\address{IPNO, CNRS/IN2P3, Université Paris-Sud, Université Paris-Saclay, F-91406 Orsay, France}}

\maketitle
\begin{abstract}
The hot rotating nuclei could be formed in the complete and incomplete fusion reaction of two heavy ions. At low bombarding energies the reaction goes via compound nucleus formation and subsequent evaporation of charged particles, neutrons and $\gamma$-rays. However, with increasing the energy of the projectile, the emission of particles during the equilibration process becomes more and more probable. This effect can be estimated by the Heavy-Ion Phase-Space Exploration (HIPSE) code which describes the production of clusters of various size from nucleons initially in the target or projectile. This dynamic evolution finalizes with the compound nuclei, quasi-fission or multi-fragmentation products. The hot rotating nuclei produced in fusion reaction can de-excitate by evaporation of particles and emission of $\gamma$-rays from the Giant Dipole Resonance, or by fission into two fragments. These processes, evaporation and fission, are described within statistical codes such as GEMINI++ or in dynamical approaches by solving the transport equations of Langevin type. In the present article we will concentrate on the possible effect of the pre-equilibrium emission on the strength function of the effective Giant Dipole Resonance, which can be described within Thermal Shape Fluctuation Model (TSFM) approach.

\end{abstract}
\PACS{24.30.-v, 25.70.Jj}
  
\section{Introduction}

The creation of the hot nucleus by fusion of two nuclei is still an open issue. The simplest idea about compound nucleus (CN) formation is to add the masses and charges of participants and to calculate the excitation energy taking into account the reaction heat. But more advanced approach is the HIPSE (Heavy-Ion Phase-Space Exploration) \cite{lacroix:2004}, an event generator describing the 
pre-equilibrium formation of clusters from nucleons of the projectile or target. Using iterative method it allows to build nuclei with various sizes, depending on the number of participating nucleons and an available energy. In practice, the outcome of the code is the ensemble of nuclei (prefragments) and particles, which mimic the transfer of few nucleons between projectile and target (quasi-projectile/quasi-target nuclei) and fusion process, where particles are emitted during the equilibration of the system. The energy and momentum conservation laws ensure correct estimation of the angular momentum and excitation energy of produced prefragments. Such an ensemble of hot and rotating nuclei subsequently de-excites either by emitting particles and high-energy $\gamma$-rays from the Giant Dipole Resonances (GDR) or by fission. 

For the GDR strength function calculation the Thermal Shape Fluctuation Model (TSFM) \cite{dubray:2005} appeared to be very successful. It allows to connect the shape of GDR function with the nuclear surface form, and it was successfully demonstrated e.g. by predicting the Jacobi shape transitions in rotating nuclei at the highest spins  \cite{kmiecik:2007}. The shape of the GDR depends on the deformation of the nucleus, thus in the TSFM GDR strength function are estimated for the full ensemble of nuclear forms and weighted within the Boltzmann factor. The free energy is obtained with the potential energy surfaces (PES) for angular momenta in the range of 2-100~$\hbar$ and the entropy are calculated with the $A_{CN}/7$ nuclear level density parameter. The PES are acquired with the Lublin-Strasbourg Drop model \cite{pomorski:2003}.

The reaction of $^{48}$Ti+$^{40}$Ca with 300 and 600~MeV beam energies has been previously analyzed with GEMINI++ \cite{charity:2010, ciemala:2013} and results were compared with existing experimental data \cite{ciemala:2015, ciemala:2017, valdre:2016}. Such reactions lead via complete fusion to the compound nucleus $^{88}$Mo, which was studied before or to the lighter systems (prefragments) after pre-equilibrium particle emission. Therefore present work will focus on the estimation of the pre-equilibrium emission influence on the de-excitation channels, and especially on influence on the effective strength function of  the GDR. The preliminary results of influence of the pre-equilibrium particle emission on the mass distribution of compound nuclei, their angular momenta and excitation energies are presented here. 

\section{Results}
\begin{figure}[htb]
\centerline{%
\includegraphics[width=9.5cm]{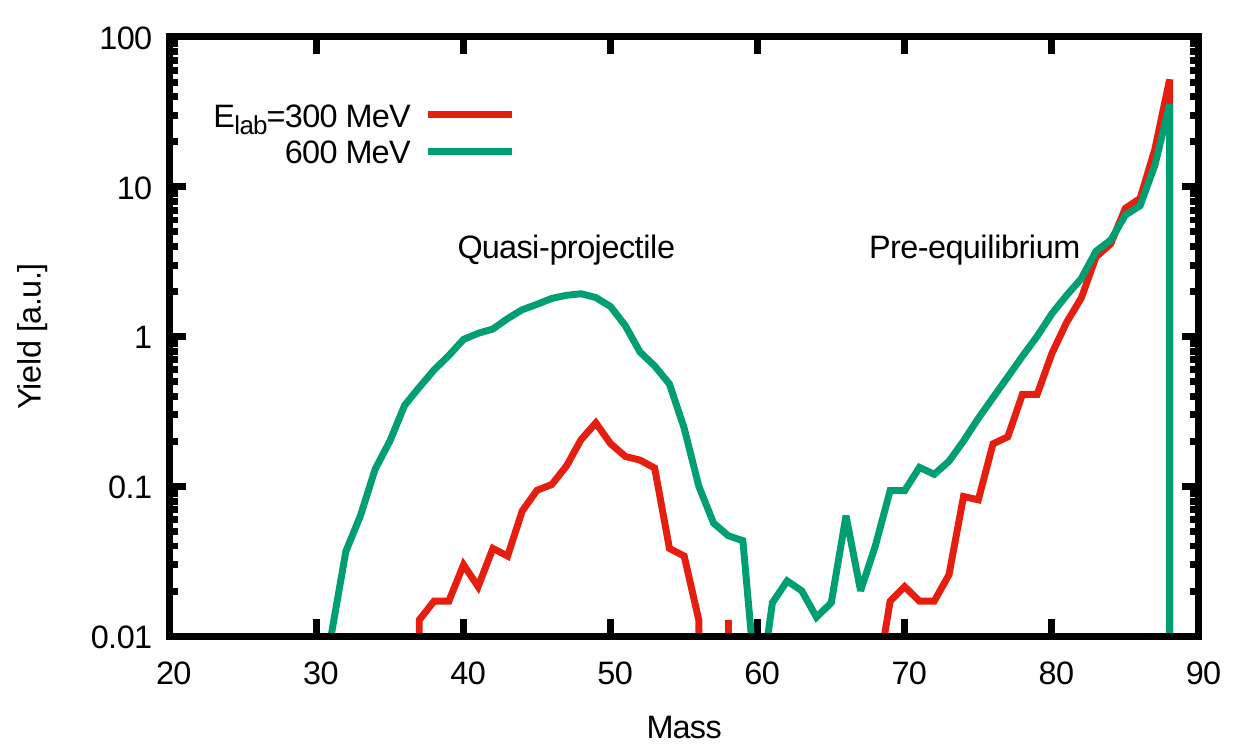}}
\caption{The mass distribution of the nuclei produced in reaction $^{48}$Ti+$^{40}$Ca with 300 and 600~MeV beam energies obtained with the HIPSE code. The quasi-projectile/quasi-target nuclei are focused around A=45, prefragments, obtained by pre-equilibrium emission, are in range 60$<$A$<$88, and finally, at A=88, are the compound nuclei.}
\label{Fig01}
\end{figure}

\begin{figure}[htb!]
\centerline{%
\includegraphics[width=9.5cm]{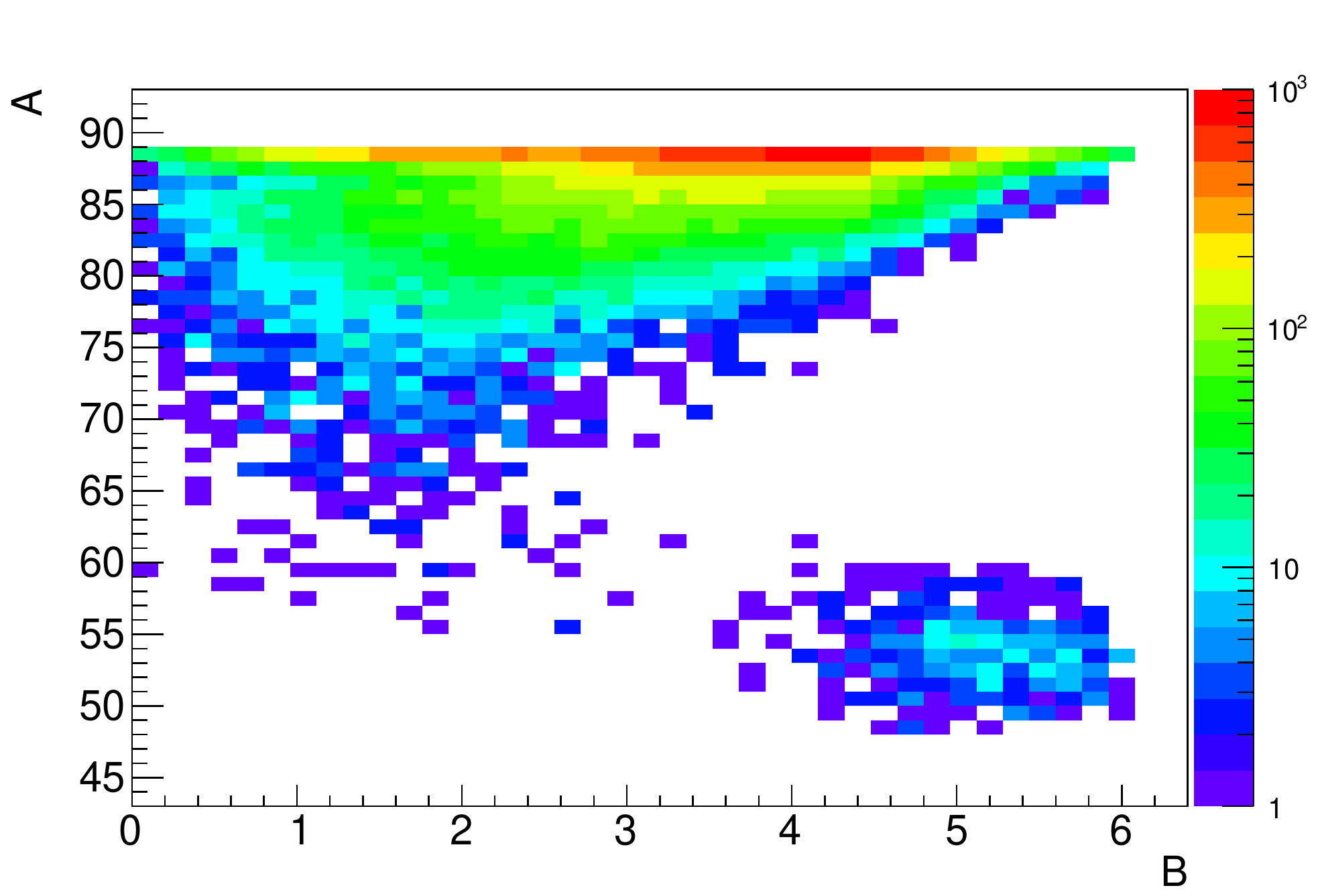}}
\caption{The correlation of the prefragment mass with the distance between the impact parameter in $^{48}$Ti(600~MeV)+$^{40}$Ca process. }
\label{Fig02}
\end{figure}

The reaction of $^{48}$Ti+$^{40}$Ca with 300 and 600~MeV beam energies lead to excitation energies of the compound nuclei 123.8 and 260.7~MeV, respectively. At such high beam energies the pre-equilibrium emission of the particles starts to play a role. The HIPSE event generator provides the full ensemble of nuclei, with their excitation energies and angular momenta, which could be created during the collision of Titanium and Calcium nuclei. Fig.~\ref{Fig01} shows the difference between distributions of nuclear masses estimated by HIPSE for two beam energies: 300 and 600~MeV. The bump around mass A=45 is connected with construction of the quasi-projectile/quasi-target nuclei by transfer reaction. This effect is very interesting but out of the scope of our work. The important part are nuclei with masses more then 60, which are produced by the emission of particles before the target-projectile system is fully equilibrated. With increasing the excitation energy the range of possible species are broader but for reaction at 600~MeV beam energy we are on the limit to clearly distinguish between quasi-projectile/quasi-target nuclei and pre-equilibrium emission part.
Furthermore, the HIPSE outcome allows to correlate the mass of the created nucleus with the distance between the center-of-mass of the projectile and the target (impact parameter), as it is displayed in Fig.~\ref{Fig02}. The large impact parameter provides two ranges of masses: 40-60 and more then 60. This confirm our statement about quasi-projectile/quasi-target production as these phenomenon is expected in the peripheral collision mostly.

\begin{figure}[htb!]
\centerline{%
\includegraphics[width=9.5cm]{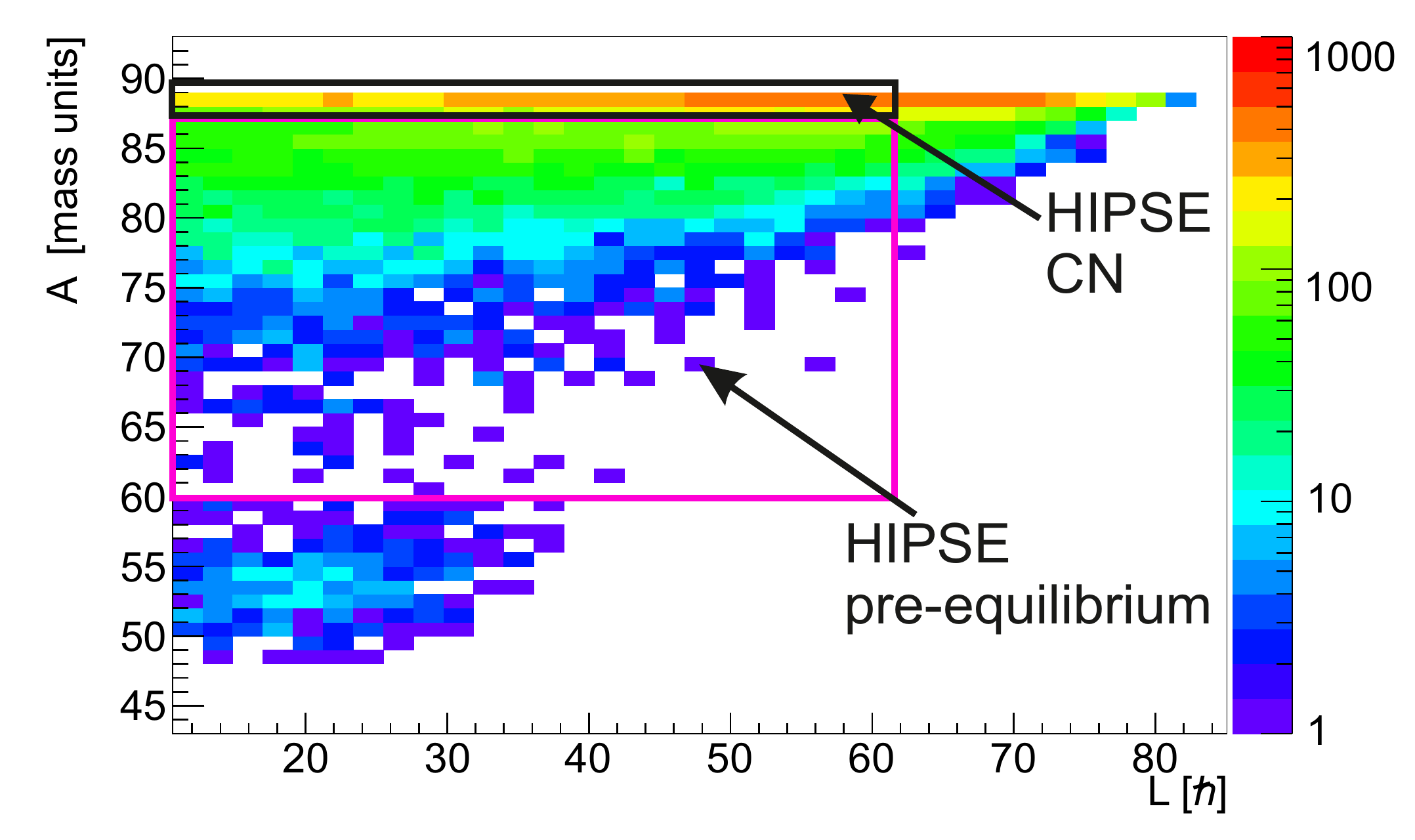}}
\caption{The mass vs. angular momentum distribution of equilibrated compound nuclei created by HIPSE code for $^{48}$Ti(600~MeV)+$^{40}$Ca reaction. The $L_{cut}$=64~$\hbar$ marks the spin at which the fission barrier vanishes. Two areas are connected with the $^{88}$Mo compound nucleus (HIPSE CN) and prefragments created with pre-equilibrium emission (HIPSE pre-equilibrium).}
\label{Fig03}
\end{figure}

\begin{figure}[htb!]
\centerline{%
\includegraphics[width=6.5cm]{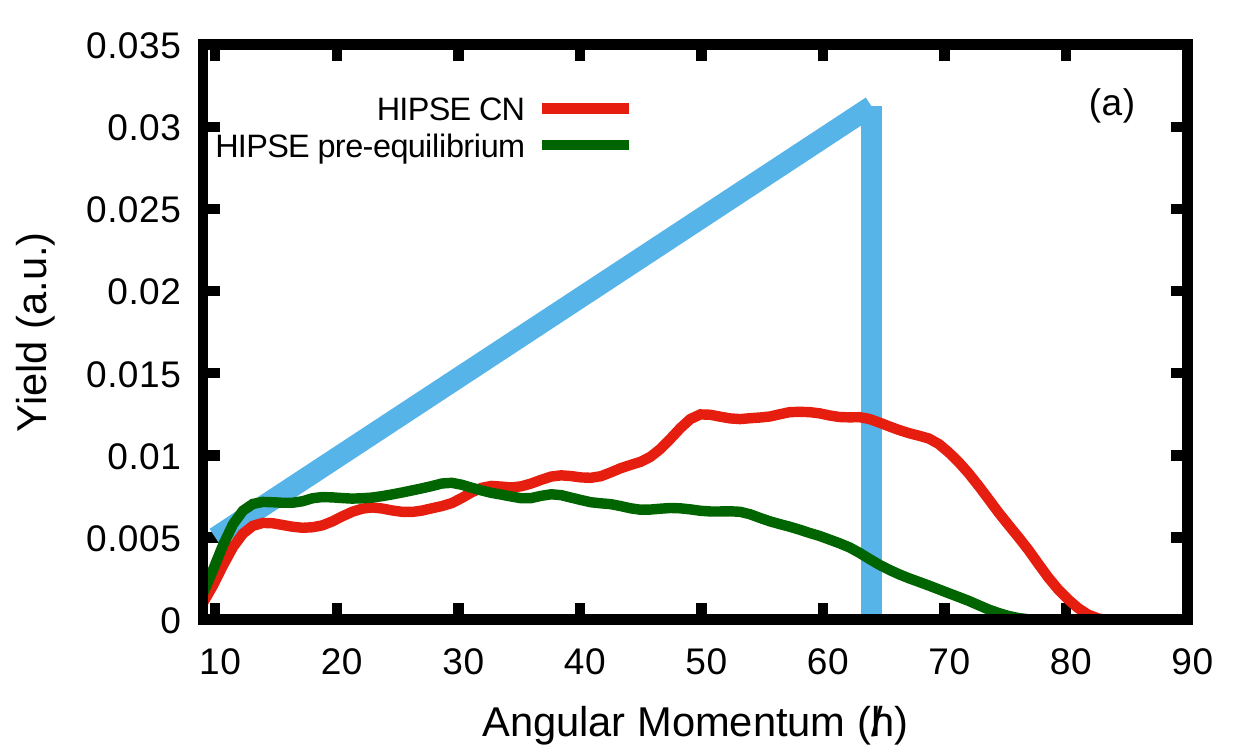}
\includegraphics[width=6.5cm]{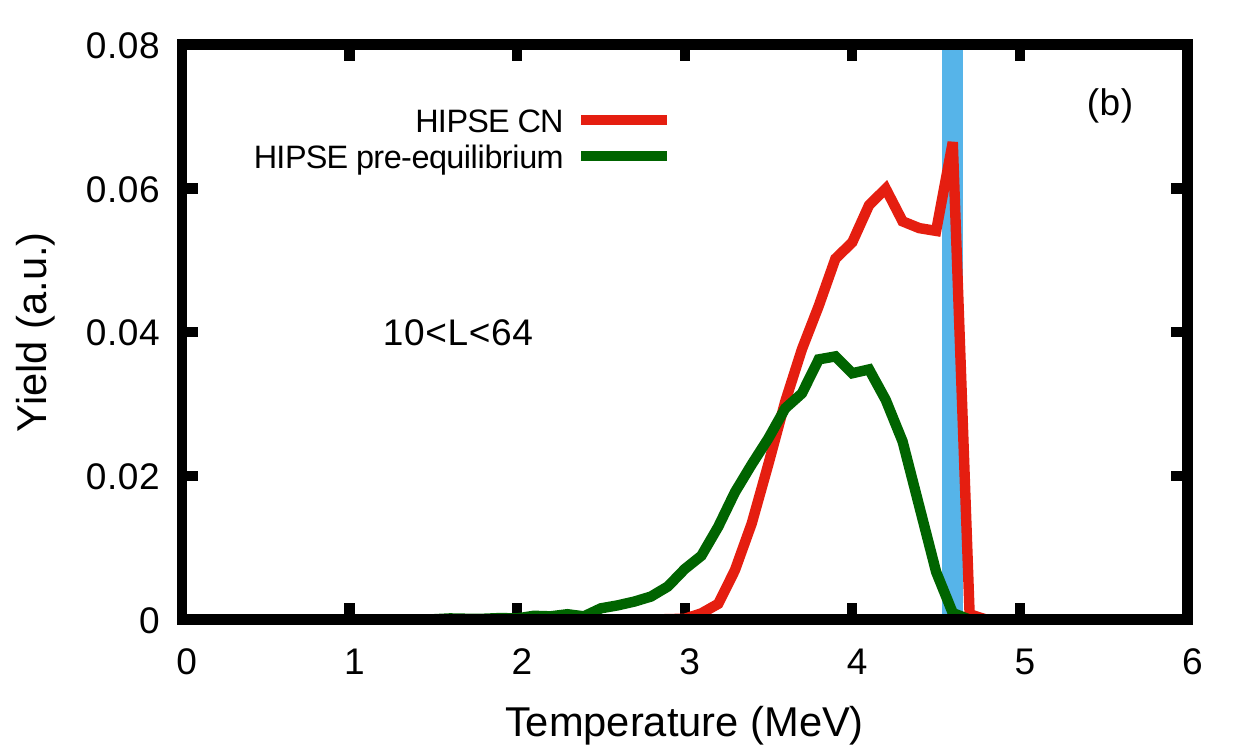}}
\caption{The angular momentum (a) and the temperature distribution (b) of equilibrated compound nuclei created in $^{48}$Ti(600~MeV)+$^{40}$Ca reaction. The thick (blue) line shows the angular momentum distribution (a) and temperature (b) taken usually for hot compound nucleus $^{88}$Mo de-excitation.}
\label{Fig05}
\end{figure}

The nuclei produced during the equilibration process may have lower masses than the compound nucleus and are characterized by distributions of angular momentum and excitation energy (usually converted to temperature). The distribution of the expected angular momentum for various prefragment masses is presented in Fig.~\ref{Fig03}. Two areas marked in the plot distinguish the events without  and with pre-equilibrium emission. First (HIPSE CN) contains only the full compound nuclei with spin and temperature marked in Fig.~\ref{Fig05} by a red curve, and the second region comprises events where masses are lower than A=88, which means that particles have been emitted before nucleus reached the equilibrium.  
The green line in Fig.~\ref{Fig05} shows the spin and temperature distributions of such prefragments. 
As a reference, the triangular distribution, assuming $l_{max}$=64~$\hbar$, is shown as a thick blue line in Fig.~\ref{Fig05} .

As fission process of light nuclei is not yet properly treated in HIPSE, the angular distribution is reaching non-realistic high spins up to 80~$\hbar$. Therefore, in the subsequent analysis the fission limit was set manually to 64~$\hbar$, as such value was estimated from the experiment \cite{ciemala:2017}. In addition, in the analysis, we will focus our attention on masses A$>$60, angular momentum range 10-64~$\hbar$ and the reaction with higher beam energy where the effects of pre-equilibrium emission should be better visible.   
\begin{figure}[htb!]
\centerline{%
\includegraphics[width=9.5cm]{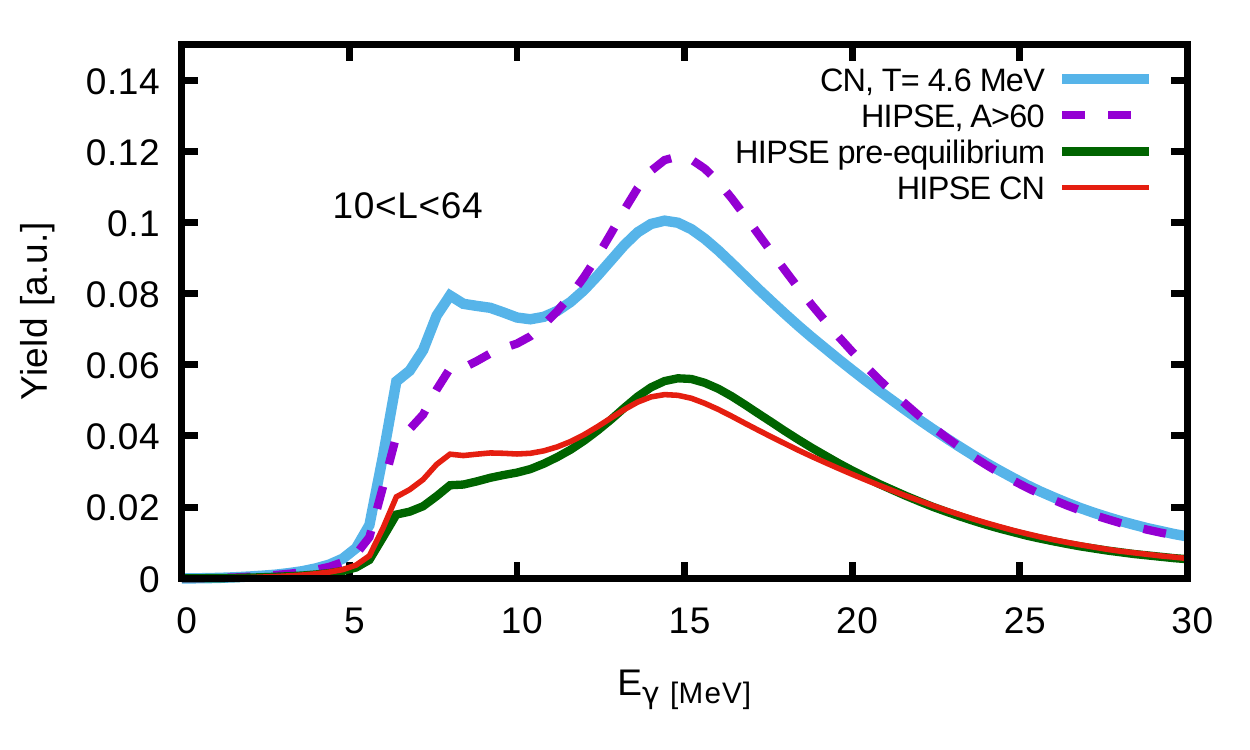}}
\caption{The strength functions of the GDR built in compound nucleus $^{88}$Mo (CN) and for ensemble of nuclei generated in the HIPSE code: the compound nucleus (HIPSE CN), the nuclei produced with pre-equilibrium emission (HIPSE pre-equilibrium) and integrated over the prefragment mass distribution (HIPSE A$>$60). All calculations were done for 10-64~$\hbar$ range of angular momentum.}
\label{Fig04}
\end{figure}
The pre-equilibrium emission process might affect the decay of equilibrated excited nuclei determining their mass, temperature and angular momentum distributions. Therefore, it is interesting to study the decay observables as emitted particles and $\gamma$-rays. Among them high-energy gamma rays from GDR decay are of importance since they can deliver information on the nuclear shape. To check the influence of pre-equilibrium emission on GDR strength function the results of the two approaches were confronted. 

In the first approach, a traditional picture assuming the $^{88}$Mo compound nucleus at the temperature of 4~MeV and triangular spin distribution (Fig.~\ref{Fig05} a,b), which de-excites by emission of high-energy $\gamma$-rays. In the second, the event generator HIPSE was used to provide more realistic data, taking into account pre-equilibrium emission. The GDR shapes calculated for nuclei obtained using these two different methods are shown in Fig.~\ref{Fig04}. The GDR strength functions were calculated using TSFM, and only first chance GDR emission was considered.

As can be seen in Fig.~\ref{Fig04}, the shapes of GDR strength functions for emission from compound nucleus are very similar for triangular angular momentum distribution (blue line) and from HIPSE (red line). Both of them show split GDR strength function with a lower energy component around 8~MeV, indicating strong component from very deformed nuclei. Contrary, the GDR strength function for GDR built in nuclei after pre-equilibrium emission (green line) has reduced low energy component, meaning that the contribution of nuclei with large deformation is smaller. This seems to be compatible with the lower mean value of the angular momentum distribution (see Fig.~\ref{Fig05}, a). The final effective GDR strength function (dashed purple line), being the sums of the two GDR strength functions, differs substantially from the one assuming triangular angular momentum distribution.

So the general message from this simple considerations is that the pre-equilibrium emission might considerably change the observed effective GDR strength function, what should be taken on account in the analysis of the experimental data.
Additional point, which was not considered in this paper, is that at such high temperatures the GDR emission during evaporation process has to be taken on account, when comparing experimental results with the theoretical predictions.

\section{Summary}
The study of the pre-equilibrium particle emission is crucial for discussion of de-excitation of hot nuclei.
The preliminary estimation of the influence of the pre-equilibrium emission on the shape of the GDR strength function has been done with the Thermal Shape Fluctuation Model. 
The difference between GDR emitted from HIPSE CN and standard CN is due to higher spin influence in the later. The pre-equilibrium emission causes the lowering of the spin and temperature of prefragments thus the low-energy component in GDR spectrum is suppressed.
 \vskip 0.5cm
{\bf Acknowledgements}\\
This work has been supported by the COPIN-IN2P3 Polish-French Collaboration 
and LEA COPIGAL.


\end{document}